\documentstyle[12pt]{article}
\font\tenbf=cmbx10
\font\tenrm=cmr10
\font\tenit=cmti10
\font\elevenbf=cmbx10 scaled\magstep 1
\font\elevenrm=cmr10 scaled\magstep 1
\font\elevenit=cmti10 scaled\magstep 1

\def\boldnabla{\mbox{\boldmath{$\nabla$}}}

\def\boldtau{\mbox{\boldmath{$\tau$}}}
\def\boldpsi{\mbox{\boldmath{$\psi$}}}
\def\Bsi{\mbox{\boldmath{$\psi$}}}
\def\Tauiii{\mbox{\boldmath{$\tau$}}_{\bf 3}}
\renewenvironment{thebibliography}[1]
 { \elevenrm
   \begin{list}{\arabic{enumi}.}
    {\usecounter{enumi} \setlength{\parsep}{0pt}
     \setlength{\itemsep}{3pt} \settowidth{\labelwidth}{#1.}
     \sloppy
    }}{\end{list}}

\begin{document}
\begin{center}%
{{\tenbf SUPERCONDUCTING PHASE TRANSITIONS\\
\vglue 10pt
IN 2+1 DIMENSIONAL QUANTUM FIELD THEORIES MODELING\\
\vglue 10pt
GENERALIZED POLARONIC INTERACTIONS\\
\vglue 20pt
I: JAHN-TELLER INSPIRED MODELS%
\footnote{Presented at
          {\it Workshop at Differential Geometry and Quantum Physics\/},
               University of Leipzig, 28.\,09.\,92 - 02.\,10.\,92,
               Leipzig, Germany.}
\\}
%\vglue 5pt
%{\ninerm (For 20\% Reduction to 8.5 $\times$ 6 in Trim Size)\\}
\vglue 1.0cm
{\tenrm RALF D.\ TSCHEUSCHNER%
\\}
\baselineskip=13pt
{\tenit I.\ Institut f\"ur Theoretische Physik
        der Universit\"at Hamburg, Jungiusstr.\ 9%
\\}
\baselineskip=12pt
{\tenit D-W-2000 Hamburg, Federal Republic of Germany%
\footnote{until 30.\,09.\,94.
          {\it Present address:\/}
          Applied Solid State Physics,
          University of Bochum, Universit\"atsstra\ss e 150,
          44780 Bochum, Germany,
          {\it email:\/}
          Ralf.D.Tscheuschner@rz.ruhr-uni-bochum.de}
\\}
%\vglue 0.3cm
%{\tenrm and\\}
%\vglue 0.3cm
%{\tenrm SECOND AUTHOR'S NAME\\}
%{\tenit Group, Company, Address, City, State ZIP/Zone, Country\\}
\vglue 0.8cm
{\tenrm ABSTRACT}}
\end{center}
\vglue 0.3cm
{\rightskip=3pc
 \leftskip=3pc
 \tenrm\baselineskip=12pt
 \noindent
We review the fundamentals of Jahn-Teller
interactions and their field theoretical modelings and
show that a 2+1 dimensional
gauge theory where the gauge field couples to \lq\lq flavored
fermions\rq\rq\ arises in a natural way from a two-band
model describing the dynamical Jahn-Teller effect.
The theory exhibits
a second order phase transition
to novel finite-temperature superconductivity.
\vglue 0.6cm}
{\elevenbf\noindent 1. Introduction}
\vglue 0.4cm
%\vglue 0.2cm
%{\elevenit\noindent 1.1. Typeset Scripts}
%\vglue 0.1cm
\baselineskip=14pt
\elevenrm
The discovery of the cuprate high-$T_C$
superconductors
\cite{Bed}
together with the fact that
there is to date no generally accepted
theory for the relevant mechanism
\cite{And},
gives physicists reasons to search for novel
scenarios responsible for macroscopic quantum
coherence phenomena in condensed matter physics.
\par
One of the main problems is the dependence
of the nature of superconducting phase transitions
on dimensionality.
Of course, one has to be careful to make statements
like \lq\lq high-$T_C$ superconductivity is essentially
a two-dimensional problem\rq\rq\ since it is not
{\elevenit a priori\/} clear how the very nature of the
superconducting phase transition in the cuprates is
{\elevenit related to\/} or {\elevenit affected by\/} some kind of
interplane coupling.
Nevertheless, there is no doubt that
a general (i.\,e.\ quantum field theoretical)
study of the possibility of
idealized two-dimensional superconductivity
at finite temperature is a challenging task.
There is an old celebrated theorem due to
Hohenberg, Mermin and Wagner, and Coleman
\cite{Hoh}
stating that {\elevenit conventional\/}
off-diagonal long range order (ODLRO)
is suppressed at any finite temperature in 2+1 dimensional
quantum systems. At this point the fundamental question of the
very existence of loopholes arises - as often in theoretical
physics.
\par
In the following we present
a proposal for a {\elevenit microscopic\/} mechanism
which may provide such a loophole and possibly explain
high-$T_C$ superconductivity or may
open the door to even more interesting superconducting materials.
It is based on an effective electron-electron (resp.\ hole-hole)
interaction induced by a generalized dynamical Jahn-Teller
effect
\cite{Brs}.
Akin to fractional statistics
this interaction is a {\elevenit phasing interaction\/},
i.e.\ one exitation {\elevenit modulates\/} the quantum
mechanical phase of the other exitation
giving rise to a net attractive interaction
for the relevant fermions.
Alternatively, a phasing interaction
may be viewed as a {\elevenit renormalization\/}
of the statistical properties of the quanta
changing their statistical identity.
It is our objective to make plausible
that it is the phasing aspect
overlooked in more conventional treatments
of the Jahn-Teller interaction
which may play a crucial role in the
mechanism of high-$T_C$
superconductivity.
\par
Our discussion is organized as follows:
We start by recalling the early history
of the Jahn-Teller theory
in molecular physics.
We proceed by reviewing the concept of the geometrical
phase (now called Berry phase and
Aharonov-Anandan phase in the
adiabatic and in the non-adiabatic cases, respectively)
\cite{Ber}
first analyzed systematically in the context
of quantum chemistry in the pioneering work
of Mead who dubbed this phenomenon
{\elevenit molecular Aharonov-Bohm effect\/}
\cite{Mea}.
Then we show how to set up a theory of strongly correlated
electrons interacting {\elevenit via\/} a
generalized dynamical Jahn-Teller interaction.
\par
Motivated by the work of Yu and Anderson
\cite{YuA}
we show that our ansatz can be expressed in terms of a
double-well tunneling event which we christened
{\elevenit solid state instanton\/}.
It is related to what is known as a {\elevenit polaron\/}
in solid state physics, but it is not exactly
the same thing.
The microscopic double-well mechanism
allows to break parity and time-reversal
{\elevenit spontaneously\/}
thus giving a {\elevenit microscopic\/} reason
for the existence of anyons without being restricted
to them.
This double-well idea is intimately related to
T. D. Lee's early 70's work on CP violation in
elementary particle physics
\cite{Lee}.
It is very amusing to observe that already
T. D. Lee relates his own concept to geometrical
ideas associated with abstract \lq\lq distortive\rq\rq\ deformations.
Notice that the actual breaking and its magnitude
of these discrete symmetries depend on a fine tuning of
the coupling parameters. The effective Lagrangean describing
such a polaronic interaction is by no means unique and
adding or discarding terms determines whether P
and T are broken or not. Moreover, there may exist the
possibility of a phase transition between an anyonic and
a Berezinski\v{i}-Kosterlitz-Thouless-type phase
\cite{BKT}
{\elevenit within\/}
the high-temperature conducting phase.
\par
Finally we show how to model the
generalized dynamical Jahn-Teller interaction
field theoretically and give an explicit
example for a quasi two-dimensional finite temperature ansatz -
firstly written down and analyzed by
Joe Kapusta and his collaborators
(without any reference to the
generalized dynamical Jahn-Teller
mechanism, however).
This model indeed shows up
a finite temperature phase transition to
novel superconductivity within a finite window
of parameters
\cite{Kap}.
Its phase structure is very reminiscent
of the phase diagram proposed by
Chakraverty in 1979 \cite{Chv}
which led Bednorz and M\"uller \cite{Bed}
to their discovery of novel superconductivity.
This model should be seen in a wider context
including two-dimensional scenarios
{\elevenit without\/} parity and time-reversal violation.
Finally, we give a heuristical argument
for macroscopic quantum coherence induced
by Jahn-Teller systems.
\par
The field theoretical ideas presented here
are based on joint work with Heinz
\cite{Hei}.
\vglue 0.6cm
{\elevenbf\noindent
        2. What is the
        Jahn-Teller effect?}
\vglue 0.4cm
As early as 1929 v.\,Neumann and Wigner
\cite{Neu}
asked a very interesting question:
Given a hermitean $n\times n$ matrix whose
entries are depending on a sequence of parameters
$\kappa_1,\kappa_2,\dots$ - how many parameters have to
be changed in order to get a collision of two
eigenvalues of the matrix. The general answer is
{\elevenit at least three\/}.
\par
Let us make this explicit for the
most simplest case $n=2$. Any hermitean
$2\times2$ matrix $A$ may be expanded
with respect to the Pauli basis:
\begin{eqnarray}
A \;=\; \alpha^\mu\sigma_\mu
  &=& \alpha^0{\bf 1} +
      \alpha^1\sigma_1 +
      \alpha^2\sigma_2 +
      \alpha^3\sigma_3    \\
  &=& \left(
      \begin{array}{cc}
      \alpha^0+ \alpha^3 & \alpha^1 +i\alpha^2\\
      \alpha^1-i\alpha^2 & \alpha^0 - \alpha^3
      \end{array}
      \right).
\end{eqnarray}
Solving the eigenvalue equation
${\rm det}(A-\lambda{\bf 1})=0$
gives
\begin{equation}
\lambda=
\alpha^0\pm
\sqrt{(\alpha^1)^2+(\alpha^2)^2+(\alpha^3)^2}.
\end{equation}
Clearly we have to send all three \lq\lq space-like\rq\rq\
$\alpha$'s to zero in order to get a collision!
\par
Evidently, in case of a real hermitean
(i.e.\ symmetric) matrix this reduces to two
parameters, in case of a real diagonal
matrix to one parameter.
\par
Interpreting the space of independent parameters
as a mechanical configuration space and cutting out
the point of conicidence, i.e.\ the {\elevenit level crossing
point\/}, we obtain manifolds with non-trivial
topological structures. This suggests that in all three
cases, once physically realized, we may expect topological
quantization effects giving rise to very interesting
physics.
\par
To get real physics from this mathematical observations
we must find a Hamiltonian having the structure
of $A$ and a mechanism which prevents the level crossing.
This was done in the pioneering work of Jahn and
Teller in 1937 who studied the stability of
polyatomic molecules in degenerate electronic states
\cite{Jah}.
\par
Essentially the famous Jahn-Teller theorem
states the following: A configuration of a polyatomic
molecule for an electronic state having orbital degeneracy
cannot be stable with respect to all displacements of the
nuclei unless the nuclei all lie on a straight line.
The original proof of Jahn and Teller
\cite{Jah}
is based on a detailed discussion of particular symmetries
and their realizations. A more general proof within the
framework of induced representations of finite groups was given by
Ruch and Sch\"onhofer in 1965
\cite{Ruc}.
It is interesting to note that -
by the side of the Weiss theory of ferromagnetism -
the Jahn-Teller effect is the archetype
of what is commonly called
{\elevenit spontaneously symmetry breaking\/} (SSB),
a fact recently recalled by Nambu
\cite{Nam}.
\par
In modern terminology a dynamics determined
by a sensible matching of {\elevenit vibr\/}ation modes
and electr{\elevenit onic\/} excitations is called a
{\elevenit vibronic interaction\/}
\cite{Brs}.
There is no doubt that a phonon induced
fermion-fermion interaction
incorporating Jahn-Teller-type effects
may exhibit new pecularities
beyond the marks of the conventional
electron-phonon interaction.
Due to the non-trivial topological structure
of the configuration spaces involved, we may
expect highly non-trivial quantization and
coherence phenomena, and it is the aim of this
discussion to push forward the thesis that the
understanding of high-$T_C$
superconductivity is {\elevenit at least\/} related to
the generalized dynamics of vibronic interactions.
\vglue 0.6cm
{\elevenbf\noindent
        3. The geometric phase in
        Jahn-Teller systems}
\vglue 0.4cm
To sum up, the (static) Jahn-Teller
effect is an electronic symmetry breaking phenomenon
associated with a spontaneous distortion.
To be concrete, let us consider a vibronic interaction
of a doubly degenerate electronic state $(E)$ with a
doubly degenerate vibrational mode $(e)$
\cite{Zwa}.
\par
According to the Jahn-Teller theorem
the nuclear motion lifts the electronic degeneracy,
i.e.\ there are nuclear configurations of lower energy
than the symmetric state.
This $E\otimes e$ Jahn-Teller effect is the
archetype of Berry's phase: A quantum-mechanical
phase shift of purely geometrical origin associated
to an adiabatic cycle starting and ending at the same
pure state
\cite{Ber}.
In our case we start by writing down
the quantum-mechanical two-dimensional
harmonic oscillator Hamiltonian
for a double degenerated vibrational mode
\begin{equation}
H = \frac{{\bf P}^2}{2M} + \frac{1}{2}M\Omega^2{\bf Q}^2,
\end{equation}
with ${\bf P}=(P_x,P_y,0)$
and ${\bf Q}=(Q_x,Q_y,0)$.
This Hamiltonian is thought to be acting on
two-component (Pauli) wave functions $\psi_i$ (i=1,2).
The vibronic coupling is given by adding a term
$k\cdot\boldtau{\bf Q}$
to $H$, whereby $\tau_i$
denote the Pauli matrices and k is a coupling constant.
(We use the letter $\tau$ to avoid confusion with spin degrees
of freedom).
In analogy to elementary particle physics
we call the associated internal quantum number
of the electron
{\elevenit a Jahn-Teller isospin\/}
or {\elevenit flavor\/}.
\par
Explicitely we have
\begin{eqnarray}
H &=& \left(
      \frac{{\bf P}^2}{2M}+\frac{1}{2}M\Omega^2{\bf Q}^2
      \right)
      \cdot{\bf 1} +
      k\cdot
      \left(
      \begin{array}{cc}
       0        & Q_x-iQ_y   \\
       Q_x+iQ_y & 0
      \end{array}
      \right)
      \\
  &=& \left(
      \frac{{\bf P}^2}{2M}+\frac{1}{2}M\Omega^2{\bf Q}^2
      \right)
      \cdot{\bf 1} +
      k\cdot
      \left(
      \begin{array}{cc}
       0                       & Q\,{\rm e}^{-i\varphi} \\
       Q\,{\rm e}^{i\varphi}   & 0
      \end{array}
      \right)
\end{eqnarray}
with $Q_x+iQ_y=Q\,{\rm e}^{i\varphi}$ and $Q_z=0$.
Let us rewrite the Hamiltonian in cylinder coordinates:
\begin{equation}
H =   \left(
      \frac{P_Q^2}{2M} +
      \frac{P_\varphi^2}{2MQ^2}+\frac{1}{2}M\Omega^2{\bf Q}^2
      \right)
      \cdot{\bf 1} +
      k\cdot
      \left(
      \begin{array}{cc}
       0                       & Q\,{\rm e}^{-i\varphi} \\
       Q\,{\rm e}^{i\varphi}   & 0
      \end{array}
      \right).
\end{equation}
In the {\elevenit adiabatic\/} or
Born-Oppenheimer {\elevenit approximation\/}
we neglegt the kinetic energy term and diagonalize
the remainder. We obtain
\begin{equation}
H_{BO}=
       \frac{1}{2}M\Omega^2{\bf Q}^2
       \cdot{\bf 1} +
       k\cdot
       \left(
       \begin{array}{cc}
       Q & 0 \\
       0& -Q
       \end{array}
       \right)
\end{equation}
with
\begin{equation}
       \left(
       \begin{array}{cc}
       Q & 0 \\
       0& -Q
       \end{array}
       \right)
=
       U
      \left(
      \begin{array}{cc}
       0                       & Q\,{\rm e}^{-i\varphi} \\
       Q\,{\rm e}^{i\varphi}   & 0
      \end{array}
      \right)
       U^\dagger.
\end{equation}
Note that the matrix
\begin{equation}
      \left(
      \begin{array}{cc}
 \displaystyle{\frac{  {\rm e}^{ i\varphi/2}}{\sqrt{2}}}  &
 \displaystyle{\frac{  {\rm e}^{-i\varphi/2}}{\sqrt{2}}}  \\
& \\
 \displaystyle{\frac{-i{\rm e}^{ i\varphi/2}}{\sqrt{2}}}  &
 \displaystyle{\frac{ i{\rm e}^{-i\varphi/2}}{\sqrt{2}}}
      \end{array}
      \right)
\end{equation}
is a {\elevenit double-valued\/} function in the polar angle
$\varphi$.
The associated energy eigenvalues are
\begin{equation}
E_\pm(Q,\varphi)=
       \frac{1}{2}M\Omega^2{\bf Q}^2\pm |{\bf Q}|,
\end{equation}
(where we have written $|{\bf Q}|$ for $Q$) and correspond to
two sheets (the upper cone-like, the lower sombrero-like)
coinciding at {\elevenit a point of degeneracy\/} at the origin.
This {\elevenit point of zero distortion\/} defines a
{\elevenit conical intersection\/}.
\par
Because of the {\elevenit double-valued\/} character
of the diagonalizing similarity transformation
the eigenstates $\eta_\pm$ corresponding to $E_\pm(Q,\varphi)$
are double-valued in $\varphi$. The multiple-valuedness may be
compensated for by an appropriate local gauge transformation
\begin{equation}
\eta_\pm\longrightarrow{\rm exp}(i\varphi/2)\,\eta_\pm,
\end{equation}
which in turn induces the change
\begin{equation}
\boldnabla\longrightarrow\boldnabla-i{\bf A},
\end{equation}
with
\begin{equation}
{\bf A}=-{\bf e}_\varphi/2Q
\end{equation}
in the nuclear energy operator.
The vector potential ${\bf A}$
corresponds to a fictitious flux tube
with strength $1/2$ confined to the origin.
In order to visualize the circuit in $\varphi$
inducing the {\elevenit quasi-spinorial\/}
sign change of the electronic states
one takes a look on a typical example
for the $E\otimes e$ Jahn-Teller
effect: the trimer. A circuit of distortions
- corresponding to the natural motions of the nuclei -
avoids the point of symmetry, at which the trimer
looks like an equilateral triangle.
\par
Of course, while it is true, that we generally assume
that the complete system must be described by a single-valued
wave function, it is the {\elevenit splitting\/} between the
{\elevenit subsystem\/} and its {\elevenit relative environment\/}
which introduces the geometrical phase factor:
Hence the multiple-valuedness of the electronic
wavefunction is compensated for by the multiple-valuedness
of the nuclear wavefunctions.
\vglue 0.6cm
{\elevenbf\noindent
4. Beyond the adiabatic approximation}
\vglue 0.4cm
In the last section we observed that in
a Jahn-Teller system (hereafter designated by JT)
the electronic wave
functions are in general multiple-valued functions
in the slow nuclear coordinates; in particular they are
double-valued in the nuclear polar angle in the case of
a trimer.
Our claim is that this could give rise to a novel
quality of an electron-electron interaction mediated
by an oscillating JT ionic configuration.
In the following - generalizing this classical
dynamical yet adiabatic JT approach
somewhat - we will go beyond the approximation of
Born and Oppenheimer and write down an ansatz
for an effective field theory
describing a dynamical {\elevenit non-adiabatic\/}
JT interaction.
\par
One reason for this proceeding lies in the fact that
e.g.\ for the case of the octahedron in the $La_2CuO_4$
superconductor we encounter {\elevenit neither\/} an appropriate
degeneracy {\elevenit nor\/} an appropriate configuration space of
distortions justifying the applicability of the \lq\lq naive\rq\rq\
JT theorem and the Mead model
described above
\cite{Cal}.
Conversely, what we expect is that the point of
degeneracy in the relevant JT-like
system is smeared-out (just like a smeared-out
Aharonov-Bohm flux line or a regularized
anyon), such that the adiabatic transformation definitely
breaks down and, in addition, the oscillating ionic arrangement
does not simply sweep through all configurations classically
possible, but tunnels between some of them instead.
We propose a scenario in which both delicate aspects
appear as natural consequences of the same fundamental
mechanism, but nevertheless, the topological quality of
the interaction, reminiscent of Meads molecular
Aharonov-Bohm effect
\cite{Mea}
will survive.
\par
To convert the nuclear  Hamiltonian
\begin{equation}
H = \frac{{\bf P}^2}{2M} + \frac{1}{2}M\Omega^2{\bf Q}^2
  + k\cdot\boldtau{\bf Q}
\end{equation}
into a field theoretical  Hamiltonian
describing an electron-electron interaction
mediated by an oscillating ionic configuration
we simply make the replacement
\begin{equation}
  k\cdot\boldtau{\bf Q}
  \;\;\longrightarrow\;\;
  -\;\sum_{{\bf k}{\bf k'}}
      (
      c^\dagger_{{\bf k'}1},
      c^\dagger_{{\bf k'}2}
      )
  \left(
  k\cdot\boldtau{\bf Q}
  \right)
  \left( \begin{array}{c}
       c_{{\bf k}1} \\
       c_{{\bf k}2}
       \end{array}
  \right),
\end{equation}
and introduce phonon field operators
such that we get a two-band field theory with
electron-phonon vertices. A more general
Hamiltonian may be obtained by introducing
weights for the different modes and adding a conventional
term. Integrating out phonons in a standard way
we obtain a four-fermion BCS-like Hamiltonian
representing interband-intraband interactions with
some \lq\lq wrong-sign\rq\rq\ couplings.
\par
Englman, Halperin, and Weger
proposed a JT theory for the
high-$T_C$ superconductivity of the
cuprates, in which the coupling between the copper
$d_\theta$ and $d_\epsilon$ states leads to a pairing
mechanism \lq\lq of the same form, but opposite sign\rq\rq\
to that of the BCS theory
\cite{Eng}.
They argued that their own ansatz is well-supported
by band structure calculations and experiments indicating
the involvement of both $d_\theta(z^2)$ and
$d_\epsilon(x^2-y^2)$ type states in the carrier states
of cuprate superconductors. Furthermore they showed
that the proposed pairing mechanism is stable against
lattice distortion even for strong coupling.
It is noteworthy to remark that
there are a number of proposals how the JT effect
comes into play in superconductors, in particular in
high-$T_C$ superconductors. Some classical papers can
be found in
Ref.\ \cite{Nes}
and more current contributions are listed in
Refs.\ \cite{Aok}.
Especially interesting is the recent work of
K. H. Johnson {\elevenit et al.}
\cite{Koh}
who argued that the observed superconductivity at 18 K
in potassium-doped fullerene is induced by a cooperative
JT coupling leading to a BCS-like
mechanism.
Topological aspects relating the JT
phenomenon and superconductivity have been almost ignored
up until now.
The only paper, to our knowledge, relating topological
quantization effects (especially fractional quantization)
to the JT effect and
superconductivity was written by Kuratsuji
\cite{Kur}.
\par
Appel pointed out to me that if
high-$T_C$ superconductivity
is due to a JT-like
scenario then the description of the relevant
mechanism surely has to go far beyond the
adiabatic approximation \cite{App}. In particular
he was inspired by the work of
Cohen and collaborators
\cite{Coh}
who emphasized that - due to the fundamental instability
of the oxygen ion which causes its motions influencing
the charge density between the copper and oxygen - anharmonic
double-well potentials for normal modes may give larger
coupling then expected from harmonic phonons and are less
sensitive to the mass. Appel argued that the
{\elevenit local phonon\/} ansatz by Yu and Anderson
\cite{YuA}
- considered as the non-adiabatic extension of the
dynamical JT effect - provides a
suitable framework to describe the fundamental interaction.
Note that we are not interested in the exact details
of the interaction
(e.g.\ apex in-plane charge interaction,
apex positional splitting,
out-of-plane motions etc.);
we only assume that the essential dynamics is
governed by an inharmonic potential.
\par
In a Yu-Anderson-type model we restrict ourselves
to a {\elevenit one\/} mode description replacing
the \lq\lq double sheeted sombrero\rq\rq\
by a \lq\lq double sheeted double well\rq\rq.
Explicitely we have
\begin{eqnarray}
H_{el-ph}  &=& -\sum_{{\bf k},{\bf k+q}}
                     (
                     c^\dagger_{{\bf k+q}s},
                     c^\dagger_{{\bf k+q}p}
                     )
\left( \begin{array}{cc}
       0
       &
       \lambda Q
       \\
       \lambda Q
       &
       0
       \end{array}
\right)
\left( \begin{array}{c}
       c_{{\bf k}s} \\
       c_{{\bf k}p}
       \end{array}
\right)\!\!.
\end{eqnarray}
To diagonalize the phonon matrix
we introduce the {\elevenit chiral\/}
- i.e. left-handed
and right-handed - linear combinations
\begin{eqnarray}
c_{{\bf k}L}
             &=&
                 \frac{1}{\sqrt{2}}
                 (c_{{\bf k}s}-c_{{\bf k}p}),\\
c_{{\bf k}R}
             &=&
                 \frac{1}{\sqrt{2}}
                 (c_{{\bf k}s}+c_{{\bf k}p}),
\end{eqnarray}
and get
\begin{eqnarray}
H_{el-ph}  &=& -\sum_{{\bf k},{\bf k+q}}
                     (
                     c^\dagger_{{\bf k+q}L},
                     c^\dagger_{{\bf k+q}R}
                     )
\left( \begin{array}{cc}
        \lambda Q
       &
       0
       \\
       0
       &
       -\lambda Q
       \end{array}
\right)
\left( \begin{array}{c}
       c_{{\bf k}L} \\
       c_{{\bf k}R}
       \end{array}
\right)\!\!,
\end{eqnarray}
i.e.\ an interaction term proportional to $\tau_3$.
Yu and Anderson proceed
by \lq\lq integrating out\rq\rq\ the electron degrees
of freedom and calculate the
dynamical modification of the
harmonic oscillator potential
giving a \lq\lq dynamical double well\rq\rq
$\propto (Q^2-h)^2$ replacing the more singular
$Q^2-|Q|$ term. The calculation is very involved
and relies heavily of path integral techniques
reminiscent of instanton
calculations in quantum field theory.
\par
An effective non-relativistic Lagrangean
for this model may have the form
\begin{equation}
{\cal L} = \left( \frac{1}{c^2}
             \partial_t\varphi
             \partial_t\varphi
            -\partial_k\varphi
             \partial_k\varphi \right)
            -V(\varphi)
            +\boldpsi^\dagger
             i\partial_t
             \boldpsi
            -\frac{1}{2m}|\partial_k
                \boldpsi|^2
            -\frac{1}{2m} \lambda  \varphi
             (\boldpsi^\dagger
              \Tauiii
              \boldpsi),
\end{equation}
where $V(\varphi)$ is a quartic term and
$\boldpsi$ is a two-component
Schr\"odinger field.
That anharmonicity modifies the mass-frequency
relation of a quantum mechanical oscillator and
hence the isotope effect is due to the non-analytic
character of the solution of the double-well tunneling
problem - a rather general feature.
In field theory tunneling events are called instantons
and it is, to our opinion, appropriate to name the
Yu-Anderson local phonon a {\elevenit solid state
instanton\/}.
Summarizing, complementary to the BCS-like
four fermion interaction
\`a la
Engelman, Halperin, and Weger
\cite{Eng}
which is obtained by integrating out the phonons,
we get an effective anharmonic phonon potential
by integrating out the fermions.
\par
A very interesting point lies in the fact that
{\elevenit double-well system\/}
is intimately related to
{\elevenit a two-level system\/}
in that the lowest states of the former are to be
identified with the only states spanning the latter.
Now the dynamics of the two-level system considered
as an abstract spin-$1/2$ system is driven by an
abstract external magnetic field - self-consistently
generated through the local phonon tunneling dynamics.
In case of a {\elevenit real\/} spin-$1/2$ system the driving
external magnetic field introduces definitely an oddness
under time reversal.
We do not really know under which conditions this oddness
under time reversal carries over to the abstract case, but
at least as a possibility it remains. The oddness under
T is also suggested by the fact that a hidden
parity violation is already present in the model due to
the interference of odd and even modes
and due to the fact that PT
should be a good symmetry in solid state physics.
\par
Hence a relativistic ansatz for a T, P, and C
invariant effective Lagrangean based on pseudoscalar
anharmonic phonons may be written
\begin{equation}
{\cal L} =   \frac{1}{2}
             (\partial_\mu\varphi)^2
            -V(\varphi)
            +\overline\psi
             (i\gamma_\mu\partial^\mu-m)\psi
            -g\cdot
             \overline\psi\gamma_5\psi\cdot\varphi,
\end{equation}
with
\begin{equation}
V(\varphi)= -\frac{1}{2}\lambda\varphi^2
            +\frac{1}{4}\kappa\varphi^4,
\end{equation}
where we demand $\lambda>0,\kappa>0$, and
the relativistic spin degrees of freedom are identified
with the two bands of the generalized JT
interaction in the non-relativistic limit.
\vglue 0.6cm
{\elevenbf\noindent
5. A phase transition towards
two-dimensional superconductivity}
\vglue 0.4cm
Our relativistic
Lagrangean is identical to the one studied
by T. D. Lee in the early 70's as a simple
example for spontaneous T violation
\cite{Lee,Hei},
a phenomenon discussed in the framework
of anyon physics
\cite{Kal}.
Unfortunately,
the experimental situation is
compatible with the absence of anyons
in high-$T_C$ materials rather than with their presence
\cite{Lyo}.
Nevertheless,
if high-$T_C$ superconductivity
is still a really two-dimensional phenomenon,
then we should continue to study
anyon superconductivity because it is
a nice \lq toy model\rq\ possibly exhibiting features
of the true theory.
Therefore our stategy is to develop a
theory in which anyons appear as a consequence
of a microscopic mechanism and induce
{\it finite temperature\/} superconductivity.
At the end we will try to find out how to
modify the model in order to preserve P and T.
\par
Assuming they exist,
anyons are never elementary particles like electrons
such that the question remains: How can we get
anyons and hence spontaneously T violation
from fundamental electronic interactions?
We think
it is near at hand and much more natural
to {\elevenit reverse\/} the standard argumentation
and consider the violation of time reversal
and not necessarily the validity of the two-dimensional
description as the main problem.
This view is also supported by
Wilczek's {\elevenit axion electrodynamics\/}
\cite{Wil}
and the {\elevenit chiron model\/} by
Chaplin and Yagamishi
\cite{Cha}.
Both models go beyond two dimensions
while preserving the P and T
violating character.
Indeed, we should find the {\elevenit reason\/} for anyons -
departing from a {\elevenit microscopic\/} picture. Our picture
is that it is a background of tunneling impurities
which does generate an effective four fermion interaction.
In particular, we expect that it should be possible
to derive anyon physics as a consequence of this ansatz
casted in the form of T. D. Lee's Lagrangean.
\par
Let us briefly sketch the original Lee mechanism.
Though the Langrangean is invariant under time
reversal and parity, the vacuum expectation value
\begin{equation}
\langle\varphi\rangle_{vac}=\varrho\not=0
\end{equation}
of the $\varphi$ field is not: It changes its sign under
P and T.
\par
In the tree graph approximation $V'(\varphi)=0$
we have $\varrho^2=\lambda/\kappa$.
Quantum fluctuations
\begin{equation}
\varphi=\varrho+\delta\varphi
\end{equation}
yield an effective potential
\begin{equation}
V=V_0+
  \frac{1}{2}\mu^2(\delta\varphi)^2+
  \kappa\varrho(\delta\varphi)^3+
  \frac{1}{4}(\delta\varphi)^4
\end{equation}
with
$V_0=-\kappa^{-1}\lambda^2/4$
and
$\mu^2=2\lambda$.
The mass of the $\delta\varphi$ quantum
does not vanish: Since T is a discrete
symmetry, we have no Goldstone modes here.
\par
To conclude,
we just have described an effective field theory
of a generalized dynamical JT effect
incorporating spontaneous T violation.
The main input are the two flavors and the double-well,
i.e.\ the breaking and restoring of a symmetry
associated to a microscopic degeneracy.
How can we derive an effective two-dimensional
theory from this picture?
\par
Evidently, there must exist a description of this
scenario in terms of a {\elevenit four-fermion interaction\/}.
In the spirit of Bjorken's work who - motivated by the BCS
theory - investigated the general possibility of constructing
a gauge field from fundamental fermionic interactions
\cite{Bjo},
we are looking for a relativistic ansatz incorporating
a non-propagating gauge coupling similar to our local phonon.
It was found by
Ogievetski\v{i} and Polubarinov (OP),
who have shown that it works
in field theories with antisymmetric tensor gauge bosons
\cite{Ogi}.
It is amusing that the latter have been called
{\elevenit phonon modes\/} by Balachandran et al.\
in an entirely different context
\cite{Ba1}.
\par
Opposed to the massive
Maxwell-Dirac Lagrangean
\begin{equation}
{\cal L}=
-\frac{1  }{4} ( \partial_\mu A_\nu - \partial_\nu A_\mu ) ^2
+\frac{m^2}{2}                A_\mu                        ^2
-\overline{\psi}(\gamma_\mu\partial^\mu+M)\psi
+ie\,\overline{\psi}\gamma_\mu\partial^\mu\psi A_\mu
\end{equation}
the most simple massive
OP phonon
quantum electrodynamics
is described by
\begin{equation}
{\cal L} =
-\frac{1}{12}
 (\partial_\mu     A_{\nu    \lambda}+
  \partial_\lambda A_{\mu    \nu    }+
  \partial_\nu     A_{\lambda\mu    })^2
+\frac{m^2}{4} A_{\mu\nu}^2
-\overline{\psi}(\gamma_\mu\partial^\mu+M)\psi
-\frac{1}{2}g\varepsilon_{\mu\nu\lambda\varrho}
 \overline{\psi}\gamma^\nu\psi\partial^\mu A^{\lambda\varrho}.
\end{equation}
Here the massless limit $m\rightarrow0$
is taken at the end of the computation.%
\footnote{In general, care must be exercised in taking
          $m\rightarrow0$ since the constraint structure
          of the underlying operator algebra changes
          discontinously in the limit.}
Since in this gauge interaction picture
the double well is no longer present,
we have to input the two flavors
by hand and obtain as the final
effective Lagangean
\begin{equation}
{\cal L} =
-\frac{1}{12}
 (\partial_\mu     A_{\nu    \lambda}+
  \partial_\lambda A_{\mu    \nu    }+
  \partial_\nu     A_{\lambda\mu    })^2
+\frac{m^2}{4} A_{\mu\nu}^2
-\overline{\Bsi}(\gamma_\mu\partial^\mu+M)\Bsi
-\frac{1}{2}g\varepsilon_{\mu\nu\lambda\varrho}
 \overline{\Bsi}\gamma^\nu\Tauiii\Bsi\partial^\mu A^{\lambda\varrho}.
\end{equation}
And now comes the point of the story:
Reducing this Lagrangean down to a
non-relativistic 2+1 dimensional situation
what is done by
cancelling the third row {\elevenit and\/} column of the
antisymmetric tensor field $A_{\mu\nu}$
and identifiying the rest with
the dual of a 2+1 dimensional vector potential $a_\mu$
such that the interaction term must have the form
\lq\lq field strength times a current
diagonal in flavor\rq\rq\ we get
a Lagrangean of the form
\begin{equation}
 {\cal L} = \frac{1}{2}
            \varepsilon^{\mu\nu\lambda}
            a_\mu \partial_\nu a_\lambda
             +\boldpsi^\dagger
              iD_t
              \boldpsi
             -\frac{1}{2m} | D_k
                 \boldpsi | ^2
             -\frac{1}{2m} f_{12}
              (\boldpsi^\dagger
               \Tauiii
               \boldpsi),
\end{equation}
where $f_{12}=\partial_1a_2-\partial_2a_1$ and
$D_\mu=\partial_\mu-iga_\mu$.
\par
Note that the coefficient of
the Chern-Simons term
is {\elevenit a priori\/} undetermined,
it is {\elevenit a posteriori\/} fixed
by the correspondence of the number of
flavors and the statistics parameter
according to Mavromatos et al.\
and others
\cite{Mav}.
The final 2+1 dimensional Lagrangean
coincides with the Lagrangean of
Kapusta et al.\ who suggested that the internal
degree of freedom may be identified with the spin,
though they did not forbid other interpretations
\cite{Kap}.
From this we get a statistical magnetic field
\begin{equation}
b=-g \; (\;\langle \psi^\dagger_R\psi_R \rangle   -
           \langle \psi^\dagger_L\psi_L \rangle\;),
\end{equation}
such that a breaking of the \lq\lq chiral\rq\rq\ invariance
in the generalized JT model
at low temperatures gives $b$ a finite value.
In a study of the finite temperature
Mei\ss ner-Ochsenfeld effect
Kapusta et al.\ arrived at a set
of four coupled integro-differential
equations, indicating that superconductivity
terminates at $T=T_C$.
With a mean field approximation and
certain values for the coupling constants
and effective mass Kapusta et al.\
arrived at reasonable $T_C$'s.
Thus Kapusta and collaborators have
shown that there exist
2+1 dimensional gauge theories
exhibiting a true second order
superconducting phase transition at finite temperature.
\vglue 0.6cm
{\elevenbf\noindent
6. Conclusions}
\vglue 0.4cm
\begin{table}
\centerline{Table 1: Possible realizations of symmetry (after Ref.\ 35)}
\vspace*{0.5cm}
\begin{tabular}{c|c|c|c|c|}
            & neutral & Wigner- &   Nambu-  & Berezinski\v{i}- \\
    & charged &  Weyl   & Goldstone &  Kosterlitz-Thouless  \\ \hline
symmetry of &         &         &           &            \\
Hamiltonian &
\raisebox{1.0ex}{$[H,Q]=0$} &
\raisebox{1.0ex}{$[H,Q]=0$} &
\raisebox{1.0ex}{$[H,Q]=0$} &
\raisebox{1.0ex}{$[H,Q]=0$} \\ \hline
symmetry of &         &         &           &            \\
ground state &
\raisebox{1.0ex}{$ Q|0\rangle = 0 $} &
\raisebox{1.0ex}{$ Q|0\rangle = 0 $} &
\raisebox{1.0ex}{$ Q|0\rangle \not= 0 $} &
\raisebox{1.0ex}{$ Q|0\rangle \not= 0 $} \\ \hline
energy gap to &       &      & no charged  & no charged    \\
charged exitations &
\raisebox{1.0ex}{infinite} &
\raisebox{1.0ex}{finite}   &
                      eigenstates           &
                                   eigenstates           \\ \hline
long wave   &         &         &           &            \\
exitations   &
\raisebox{1.0ex}{none}           &
\raisebox{1.0ex}{none}           &
\raisebox{1.0ex}{NG bosons}      &
\raisebox{1.0ex}{BKT bosons}     \\ \hline
local       &         &         &           &            \\
order parameter &
\raisebox{1.0ex}{0}             &
\raisebox{1.0ex}{0}             &
\raisebox{1.0ex}{$\not = 0$}    &
\raisebox{1.0ex}{0}             \\ \hline
correlator of &       &         &           &            \\
order field   &
              0       &
                 $e^{-rm}$              &
                                 $const$    &
                                            $r^{-\alpha}$      \\
as $r\longrightarrow\infty$ &   &   &   &   \\ \hline
\end{tabular}
\vspace*{0.5cm}
\end{table}
Let us take a look at the familiar
conventional 3+1 dimensional superconducting
phase transition from a quantum field theoretical
point of view:
The electric local gauge symmetry is
\lq\lq spontaneously broken\rq\rq\ in the
superconducting phase giving rise to a would-be
Goldstone boson absorbed into the Mei\ss ner-Ochsenfeld effect.
(We use the quotation marks indicating that from a
rigorous point of view local gauge symmentries are
never spontaneously broken according to the
Elitzur-L\"uscher theorem \cite{Eli}.)
Complementarily, we may view the transition
from the superconducting state to
the normal state as a spontaneously symmetry breaking
of a magnetic gauge symmetry whose generator is a
magnetic charge quantum number (\lq\lq vorticity\rq\rq)
carried by fictitious infinite long fluxlines
of infinite energy \cite{Ko1}.
In this framework the Goldstone bosons are the
photons or, physically speaking,
they manifest themselves
as the absence of the
Mei\ss ner-Ochsenfeld effect \cite{Ko1}.
\par
Conversely, in two space dimensions
conventional superconductivity does not
exist at any finite temperature because
fluctuations overcome energy in destroying
off-diagonal long range order. This is
the essential conclusion of the
Hohenberg-Mermin-Wagner-Coleman theorem \cite{Hoh},
and one reason to consider anyon superconductivity
was the possibility of an evasion of this theorem.
There might exist other ways to evade the assumptions
underlying this theorem and according our table
borrowed from Ref.\ \cite{Ko1} the
Berezinski\v{i}-Kosterlitz-Thouless transition is an obvious
choice \cite{BKT}.
Kovner and Rosenstein motivated this choice by the
\lq\lq growing body of experimental data
that points to the KT nature of the
superconducting phase transition in
$CuO$ materials\rq\rq\ \cite{Ko2}.
They propose a Lagrangean in which a
doublet of two-component complex
Dirac spinors couples to a vector field.
In this theory the electric gauge symmetry
is implemented in a BKT mode and the vector field
reminscent of our OP phonon degree of freedom
represents the corresponding BKT boson.
\par
Contrary to the case of
thin superconducting metal films
where we observe an vortex-antivortex
unbinding transition \cite{Heb}
the Kovner-Rosenstein vortices
are charged. This is very reminiscent of anyonic
superconductivity where vortices and quasi-particles
are one and the same entities.
From a conservative point of view,
one may argue that BKT exitations
tend to {\it suppress two-dimensional superconductivity\/}
and this suppression will be probably damped by
interlayer interactions
{\it enhancing three-dimensional superconductivity\/}.
In the case of the
Kovner-Rosenstein vortices the interplane
coupling shifts the values obtained in a two-dimensional
theory only by a certain amount \cite{Ko2}
thus preserving the overall two-dimensional character.
This seems to be compatible with our
JT anyon \lq toy model\rq\ \cite{Hei}.
\par
We think there is still a lot of experimental work
to be done
to find out what is really happening microscopically.
But there is one point which cannot be overemphasized:
It is by no means sufficient to confine ourselves
to a discussion of the critical behavior of the
systems we are interested in. Critical phenomena
encode a wide category of specifications into
the same universality class (e.g.\ \lq 3D XY\rq).
Critical behavior
is distinguished by the equal importance
of all length and time scales at a certain point
in the phase diagram.
Thus there is no logic which allows
us to deduce statements about
the microscopic mechanism {\elevenit from\/} critical
behavior \cite{Sch}.
Even the dimensionality and the internal symmetry
of a system cannot be read off from the critical behavior
of a system (e.g.\ asymptotic freedom of two dimensional sigma models
vs.\ asymptotic freedom four dimensional Yang-Mills theories).
Expressed in different words, the question whether
the superconducting phase transition in the cuprates
is of a novel type cannot be answered from the study
of critical indices alone.
\par
In conclusion, we have formulated a theoretical model,
in which the fermions
are interacting {\elevenit via\/} a generalized
JT interaction
leading to a superconducting phase transition
violating parity and time reversal.
It is an example for a quantum mechanical
distinction between left- and right-handedness
different from other interesting proposals doing this
\cite{Ba2}.
Our model may be regarded as a
variant of a more general BKT mechanism
for two-dimensional superconductivity
including scenarios without P and T violation.
Moreover, a phase transition between
a T preserving and T violating phase
{\it within\/} the superconducting state
seems to be possible.
%
%  \par
%  An electronic environment living on
%  a background of tunneling impurities
%  may exhibit a novel form of
%  macroscopic coherence.
%  In some sense our idea may be regarded as
%  the inversion of the solution
%  of a famous paradox formulated by Hund
%  who asked the question why one
%  does not encounter interference states
%  on the macroscopic level
%  (e.g. you here {\elevenit and\/} you there)
%  \cite{Hun}.
%  The answer is: Because of quantum damping.
%  A two level system coupled to an incoherent environment
%  will immediately fall into one of the states, say left
%  or right. But if the environment coupled to
%  the two-level system is coherent, say superconducting,
%  the tunneling dynamics of the two level system must
%  self-consistently dominate.
%
\vglue 0.6cm
{\elevenbf\noindent
7. Acknowledgements}
\vglue 0.4cm
I would like to thank
J.\,Appel,
C.\,Bandte,
M.E.\,Carrington,
A.\,Heinz,\linebreak
J.\,Kapusta,
C.A.\,Mead,
K.A.\,M\"uller,
Y.\,Nambu,
R.L.\,Stuller,
and M.\,Weger
for useful discussions.
Thanks also
to Professor K.A.\,M\"uller
for drawing my attention onto Ref.\ \cite{Nes,Sch}
and to Professor R.\,Haag for calling
my attention to Ref.\ \cite{Ruc}.
Financial support by IBM is also
greatly appreciated.
\vglue 0.6cm
{\elevenbf\noindent
8. References}
\vglue 0.4cm

\end{document}